\documentclass[epj,referee]{svjour}
\usepackage{graphics}

\begin{document}
\title{Absolute resonance strengths in the $^{6,7}$Li($\alpha,\gamma$)$^{10,11}$B reactions}
\author{Gy. Gy\"urky\inst{1}\thanks{\emph{corresponding author, 
E-mail: gyurky@atomki.hu}} \and Zs. F\"ul\"op\inst{1} \and E. Somorjai\inst{1} \and G. Kiss\inst{1,2} \and C. Rolfs\inst{3}
}                     
\institute{Institute of Nuclear Research (ATOMKI), H-4001 Debrecen, P.O.Box 51, Hungary 
\and University of Debrecen, H-4010 Debrecen, Egyetem t\'er 1., Hungary \and Institut f\"ur Physik mit Ionenstrahlen, Ruhr-Universit\"at Bochum, Bochum, Germany}
\date{Received: date / Revised version: date}
\abstract{The absolute  strengths of the E$_{\alpha}$\,=\,1175\,keV resonance in the $^6$Li($\alpha,\gamma$)$^{10}$B reaction and of the E$_{\alpha}$\,=\,814\,keV resonance in the $^7$Li($\alpha,\gamma$)$^{11}$B reaction have been measured to
$\omega\gamma$\,=\,366\,$\pm$\,38\,meV and $\omega\gamma$\,=\,300\,$\pm$\,32\,meV, respectively, in good agreement with previous values.
 These resonances can be used to measure the absolute acceptance of the recoil separator ERNA to a precision of about 10\%.  
} 
\maketitle
\section{Introduction}
\label{intro}

The $^{12}$C($\alpha,\gamma$)$^{16}$O reaction (Q\,=\,7.16\,MeV) is one of the key reactions of nuclear astrophysics \cite{rolfs}. For this reason its cross section at the relevant Gamow energy of E$_0$\,=\,0.3\,MeV must be
known with a precision of at least 10\%. In spite of tremendous experimental efforts over the last thirty years one
is still far from this goal. 
To improve the situation, a new experimental approach is in preparation at the 4\,MV Dynamitron tandem accelerator in Bochum, called ERNA -- European Recoil separator for Nuclear Astrophysics \cite{rogalla}. In this approach, the reaction is studied in inverse kinematics, $^4$He($^{12}$C,$\gamma$)$^{16}$O, {\it i.e.} a $^{12}$C ion beam is guided into a windowless $^4$He gas target and the $^{16}$O recoils are counted in a $\Delta$E-E telescope placed in the beam
line at the end of the separator, which filters the intense $^{12}$C projectiles from the $^{16}$O
recoils. 
One of the most important separator characteristics is its acceptance in angle and energy. Due to the emission of the capture $\gamma$-rays, the recoils emerging from the gas target have sizable angular and energy spreads. In order to make a reliable cross section measurement, ERNA must transport the
$^{16}$O recoils (of chosen charge state) with 100\% transmission from the gas target to the telescope.

\begin{table*}
\caption{Results of previous strength measurements.}
\label{prevtab}
\begin{tabular}{lc@{\extracolsep{2cm}}lc}
\hline\noalign{\smallskip}
\multicolumn{2}{c}{$^6$Li($\alpha,\gamma$)$^{10}$B} & \multicolumn{2}{c}{$^7$Li($\alpha,\gamma$)$^{11}$B} \\
\multicolumn{2}{c}{E$_{\rm R,c.m.}$\,=\,706\,keV, $\Gamma_{\rm R}$\,=\,1.7\,eV}  & \multicolumn{2}{c}
{E$_{\rm R,c.m.}$\,=\,518\,keV, $\Gamma_{\rm R}$\,=\,1.8\,eV} \\
author & $\omega\gamma$ [meV] & author & $\omega\gamma$ [meV]   \\
\noalign{\smallskip}\hline\noalign{\smallskip}
Meyer-Schutzmeister and Hanna \cite{meyer}& 330$\pm$80 & Bennett {\it et al.} \cite{bennett}& 600$^a$\\
Alburger {\it et al.} \cite{alburger}& 410$\pm$90 & Jones {\it et al.} \cite{jones}& 630$^a$ \\
Forsyth {\it et al.} \cite{forsyth}& 440$\pm$70 & Green {\it et al.}  \cite{green}& 500$^a$ \\
Spear {\it et al.} \cite{spear}& 400$\pm$60 & Hardie {\it et al.} \cite{hardie}& 310$\pm$47 \\
adopted value \cite{ajzenberg1}& 400$\pm$40 & adopted value \cite{ajzenberg2}& 310$\pm$47 \\
\hline
$^a$ uncertainty not given
\end{tabular}
\end{table*}

One way of determining the acceptance experimentally is to use nuclear reactions with well known absolute cross sections, such as resonant $\alpha$-capture reactions involving the $^4$He gas target of ERNA. If a reaction with a chosen resonance energy has similar kinematics as $^4$He($^{12}$C,$\gamma$)$^{16}$O, the angular and energy spreads of the $^{16}$O recoils can be simulated. This condition is fulfilled by the narrow and strong resonances at E$_{\rm R,c.m.}$\,=\,706\,keV in 
$^4$He($^6$Li,$\gamma$)$^{10}$B (Q\,=\,4.46\,MeV, $\Gamma_{\rm R}$\,=\,1.7\,eV) and E$_{\rm R,c.m.}$\,=\,518\,keV in \linebreak
$^4$He($^7$Li,$\gamma$)$^{11}$B (Q\,=\,8.66\,MeV, $\Gamma_{\rm R}$\,=\,1.8\,eV). However, the strength $\omega\gamma$
of these resonances is not known with sufficient precision, {\it i.e.} better than 10\% (table \ref{prevtab}).
Thus, we remeasured both strengths.

\begin{figure}
\hspace{0.15cm}
\resizebox{0.95\columnwidth}{!}{%
  \rotatebox{270}{\includegraphics{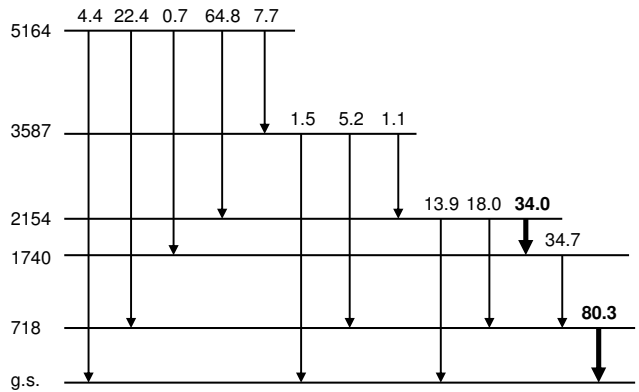}}
}
\caption{Decay scheme of the E$_{\rm x}$\,=\,5164\,keV level in $^{10}$B. The relative intensities (branchings) of the different transitions - taken from Table 10.6 of Ref. \cite{ajzenberg1} - are also indicated.}
\label{6lilevel}       
\end{figure}

\section{Experimental procedure}
\label{exp}

The experiments were carried out at the 5\,MV Van de Graaff accelerator of the ATOMKI, Debrecen, Hungary, with an energy stability and energy spread of about 1\,keV. The resonance strengths were measured using an $\alpha$ beam (about 5 and 10\,$\mu$A) on Li targets. The targets were prepared by evaporating LiF with natural isotopic abundances onto 0.2\,mm thick Ta backings. The thickness of the targets  was estimated from the evaporation procedure to be about 150\,$\mu$g/cm$^2$,  consequently the $\alpha$ energy loss in the target ($\sim$200\,keV at E$_\alpha$\,=\,1175\,keV) is much larger than the widths of the resonances and the energy spread of the $\alpha$-beam. 
 The collected charge was measured with a precision current integrator. A suppression voltage was applied at the entrance of the target chamber to minimize the effects of secondary electrons. The target was cooled by air flow. The $\gamma$-ray  yield was observed with a 40\% High Purity Germanium detector. This detector was placed at 55$^\circ$ with respect to the beam direction, with the front face of the detector at a distance of 3.5\,cm from the target.
At $\gamma$-energies below 4\,MeV (relevant to $^6$Li($\alpha,\gamma$)$^{10}$B), the absolute efficiency of the detector was measured using calibrated radioactive sources. In this energy region the measured points were fitted with a $\eta=a\times E^b$ function, where $E$ is the $\gamma$-energy and $a$ and $b$ are constants. At $\gamma$-energies above 4 MeV (relevant to $^7$Li($\alpha,\gamma$)$^{11}$B) resonant reactions emitting cascade $\gamma$-rays were used to determine the efficiency: $^{27}$Al(p,$\gamma$)$^{28}$Si (at E$_{\rm R, lab}$\,=\,767 and 992\,keV), $^{23}$Na(p,$\gamma$)$^{24}$Mg (E$_{\rm R,lab}$\,=\,1318 and 1417\,keV) and $^{39}$K(p,$\gamma$)$^{40}$Ca (E$_{\rm R,lab}$\,=\,1344\,keV). \linebreak When the low energy member of the emitted cascade fell into the energy region of the radioactive sources, the efficiency was matched to the above function. The measured high energy points were fitted with a 3rd degree polynomial matching the low energy function at about 3~MeV.

\subsection{The $^6$Li($\alpha,\gamma$)$^{10}$B reaction}
\label{6li}

\begin{figure}
\resizebox{\columnwidth}{!}{%
  \rotatebox{270}{\includegraphics{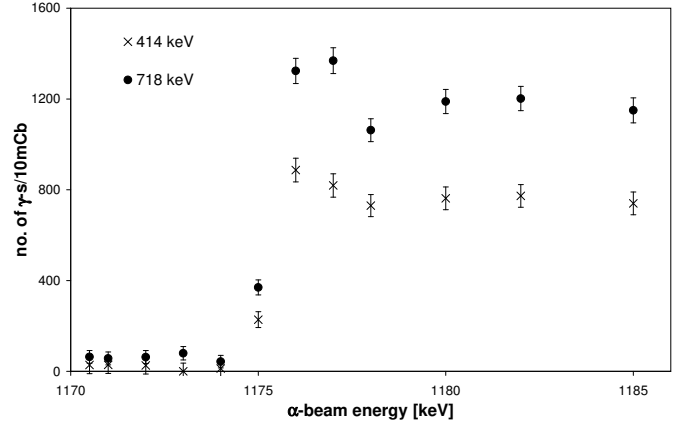}}
}
\caption{Resonance profile of the E$_{\rm R,lab}$\,=\,1175\,keV resonance in $^6$Li($\alpha,\gamma$)$^{10}$B derived from the yield of the E$_\gamma$\,=\,414 and 718\,keV lines.}
\label{6liprofile}       
\end{figure}

\begin{figure}
\resizebox{\columnwidth}{!}{%
  \rotatebox{270}{\includegraphics{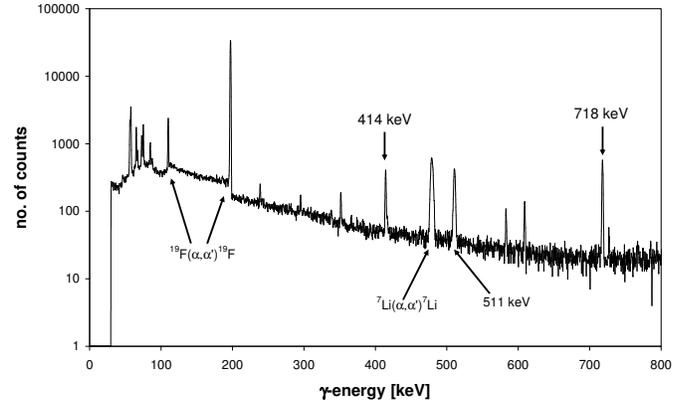}}
}
\caption{Sample $\gamma$-spectrum taken on top of the 1175\,keV resonance in $^6$Li($\alpha,\gamma$)$^{10}$B. The 414 and 718\,keV $\gamma$-lines used for the analysis as well as some other strong lines are identified.}
\label{6lispec}       
\end{figure}

The E$_{\rm R,c.m.}$\,=\,706\,keV (E$_{\rm R,lab}$\,=\,1175\,keV) resonance in \linebreak $^6$Li($\alpha,\gamma$)$^{10}$B populates the 5164\,keV excited state in $^{10}$B. Fig.\,\ref{6lilevel} shows how this state decays to the ground state. 
The most intensive cascade transitions at E$_\gamma$\,=\,414 and 718\,keV (Figs. \ref{6lilevel} and \ref{6lispec}) were used for the strength determination.
We scanned the resonance energy region at E$_{\rm R,lab}$\,=\,1175\,keV in 1\,keV steps in order to measure the resonance profile. On top of the resonance we used larger energy steps (2 to 3\,keV) in order to prove that the target is sufficiently thick, {\it i.e.} there is a horizontal plateau at energies above E$_{\rm R,lab}$. Figure \ref{6liprofile} shows the resonance profile measured with the 414 and 718\,keV $\gamma$-lines. 

\subsection{The $^7$Li($\alpha,\gamma$)$^{11}$B reaction}
\label{7li}

\begin{figure}
\resizebox{\columnwidth}{!}{%
  \rotatebox{270}{\includegraphics{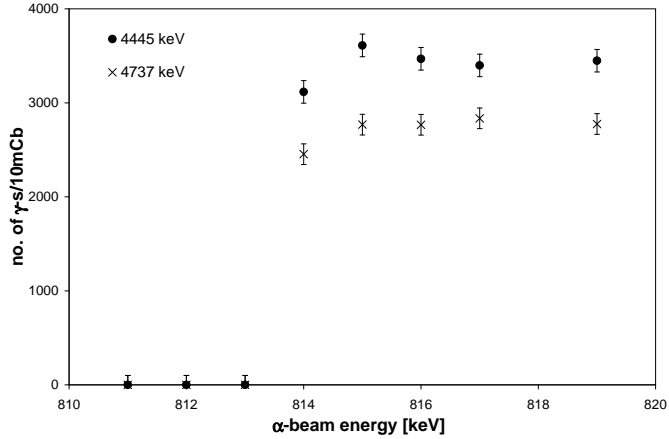}}
}
\caption{Resonance profile of the E$_{\rm R,lab}$\,=\,814\,keV resonance in $^7$Li($\alpha,\gamma$)$^{11}$B derived from the yield of the 4445 and 4737\,keV $\gamma$-lines.}
\label{7liprofile}       
\end{figure}

The E$_{\rm R,c.m.}$\,=\,518\,keV (E$_{\rm R,lab}$\,=\,814\,keV) resonance in \linebreak $^7$Li($\alpha,\gamma$)$^{11}$B populates the E$_{\rm x}$\,=\,9182\,keV excited state in $^{11}$B which cascades predominantly (86.6\%) via the \linebreak 4.44\,MeV excited state to the ground state emitting an E$_\gamma$\,=\,4737\,keV primary and an E$_\gamma$\,=\,4445\,keV secondary $\gamma$-ray. The 4445\,keV line has a small contribution (12.5\%) also from the deexcitation through the E$_{\rm x}$\,=\,6743\,keV level. These branching ratios are taken from Table 11.4 of Ref. \cite{ajzenberg2}. The resonance strength was derived from the yield of these two $\gamma$-radiations. 
Figure \ref{7liprofile} shows the resonance profile measured with the two $\gamma$-radiations. As can be seen, the requirement of a thick target is fulfilled also in this case.

\begin{figure}
\resizebox{\columnwidth}{!}{%
  \rotatebox{270}{\includegraphics{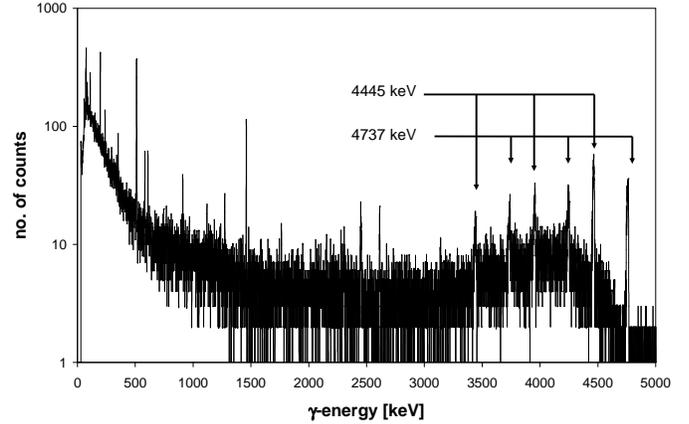}}
}
\caption{Sample $\gamma$-spectrum taken on top of the 814\,keV resonance in $^7$Li($\alpha,\gamma$)$^{11}$B. The 4445 and 4437\,keV $\gamma$-lines used for the analysis are indicated together with their escape peaks.}
\label{7lispec}       
\end{figure}

Figure \ref{7lispec} shows a typical $\gamma$-spectrum measured on top of the resonance. The E$_\gamma$\,=\,4445 and 4737\,keV lines together with their escape peaks are indicated by arrows.


\section{Analysis and results}
\label{analysis}

Using the thick target yield $y$, the resonance strength $\omega\gamma$ can be calculated using the formula
\begin{equation}
\omega\gamma=\frac{M_{\rm T}}{M_{\rm P}+M_{\rm T}}\frac{\epsilon}{2\pi^2\lambda^2}y
\end{equation}
where $M_{\rm T}$ and $M_{\rm P}$ are the masses of target and projectile, respectively, $\epsilon$ is the stopping power of the target at the resonance energy, and $\lambda$ is the de Broglie wavelength of the projectile.
The stopping power was obtained from the SRIM code \cite{SRIM}.
The thick target yield is derived from the step in the resonance function, {\it i.e.} the difference of the number of detected $\gamma$-s above and below the resonance divided by the number of incident particles and corrected for the branching ratio of the used transition and for the detector efficiency. Because of the relatively close target-detector geometry and of the $\gamma$-rays emitted in cascade transitions, summing corrections (about 10\% and 5\% for
$^6$Li($\alpha,\gamma$)$^{10}$B and $^7$Li($\alpha,\gamma$)$^{11}$B, respectively) were necessary in the case of both resonances \cite{debertin}. 

Experimental angular distributions are available in the literature for all four studied transitions in the two reactions \cite{meyer,green,warburton}. The results show that the gamma yield measured at 55 degrees represents the angle integrated yield with better than 1\% accuracy in the studied four transitions. This together with the error arising from the finite detector size has been incorported in the 7\% uncertainty of the detector efficiency.

\subsection{Results for the $^6$Li($\alpha,\gamma$)$^{10}$B reaction}
\label{6lires}

The strength of the E$_{\rm R,lab}$\,=\,1175\,keV resonance was determined from the $\gamma$-yields of the E$_\gamma$\,=\,414 and 718\,keV transitions: $\omega\gamma_{414}$\,=\,371\,$\pm$\,29\,meV and $\omega\gamma_{718}$= \linebreak 362\,$\pm$\,22\,meV. The quoted errors contain only the uncertainties of the thick target yield determination and of the reported branching ratio of a given transition. These uncertainties are specific for the two independent analyses. The weighted average of these two values is $\omega\gamma$\,= 366\,$\pm$\,17\,meV. To give the final uncertainty, systematic errors which are common for both analyses have to be included. These are from the stopping power (5\%), current integration (3\%) and detector efficiency (7\%). Adding these components quadratically, we get our final result \linebreak {\bf $\omega\gamma$\,=\,366\,$\pm$\,38\,meV}.

\subsection{Results for the $^7$Li($\alpha,\gamma$)$^{11}$B reaction}
\label{7lires}

Similarly, the strength of the  E$_{\rm R,lab}$\,=\,814\,keV resonance was determined using the yields of the 4445 and 4737\,keV $\gamma$-lines: $\omega\gamma_{4445}$\,=\,316\,$\pm$\,18\,meV and $\omega\gamma_{4737}$\,=\,285\,$\pm$\,16\,meV. The weighted average including systematic errors leads to {\bf $\omega\gamma$\,=\,300\,$\pm$\,32\,meV}.

\section{Conclusions}
\label{conc}

Our result for the strength of the E$_\alpha$\,=\,1175\,keV resonance in $^6$Li($\alpha,\gamma$)$^{10}$B 
is within the uncertainty range of the adopted value  $\omega\gamma_{\rm adopted}$\,=\,400\,$\pm$\,40\,meV \cite{ajzenberg1}. Based on the weighted average of the results of all available measurements we propose a new standard value \linebreak
{\bf $\omega\gamma_{\rm new}$\,=\,387\,$\pm$\,27\,meV} which is consistent with all measurements but its uncertainty is reduced from 10\% to 7\%.
Our new value for the strength of the E$_\alpha$\,=\,814\,keV resonance in  $^7$Li($\alpha,\gamma$)$^{11}$B
is in excellent agreement with the adopted value $\omega\gamma_{\rm adopted}$\,=\,310\,$\pm$\,47\,meV based on a single measurement. Averaging these two values we propose as a new standard value {\bf $\omega\gamma_{\rm new}$\,=\,304\,$\pm$\,26\,meV}, 
where the uncertainty is reduced from 15\% to 9\%.

With the results from previous and present work the precision of the strength of the two resonances has become better than 10\% allowing for the ERNA project (and other recoil separators) to perform a reliable acceptance measurement based on these resonant reactions.

\section*{Acknowledgments}
This work was supported by OTKA (T034259, T042733, F043408) and by DFG (436UNG113-146). Gy.\,Gy\"urky is a Bolyai fellow.

\end{document}